\documentclass[a4paper,11pt]{article}
\pdfoutput=1
\usepackage[T1]{fontenc}
\usepackage{eurosym}
\usepackage{lastpage}
\usepackage{amsfonts}
\usepackage{bbm}
\usepackage{cite}
\usepackage{xspace}
\usepackage[margin=20mm,includehead,includefoot]{geometry}
\usepackage{fancyhdr}
\usepackage{booktabs}
\usepackage{graphicx}
\usepackage{multirow}
\usepackage{array}
\usepackage{xcolor}
\usepackage{csquotes}
\usepackage{pgfgantt}
\usepackage{slashed}
\usepackage{rotating}
\usepackage[graphicx]{realboxes}
\usepackage{hyperref}
\usepackage{amsmath,amssymb}
\usepackage[normalem]{ulem}
\bibliography{bibliography}

\newcolumntype{M}[1]{>{\centering\arraybackslash}m{#1}}
\newcolumntype{N}{@{}m{0pt}@{}}
\newcommand{\be}{\begin{equation}}
\newcommand{\ee}{\end{equation}}
\newcommand{\ba}{\begin{array}}
\newcommand{\ea}{\end{array}}
\newcommand{\bea}{\begin{eqnarray}}
\newcommand{\eea}{\end{eqnarray}}
\newcommand{\p}{\partial}

\newcommand{\nn}{\nonumber}

\newcommand{\g}{\gamma}
\newcommand{\m}{\mu}

\usepackage{pgfgantt}
\usepackage{xcolor}
\usepackage[utf8]{inputenc}

\definecolor{barblue}{RGB}{153,204,254}
\definecolor{groupblue}{RGB}{51,102,254}
\definecolor{linkred}{RGB}{165,0,33} 

\usepackage{xcolor,colortbl}
\begin{document}
\baselineskip 24pt

\begin{center}
{\Large \bf Some Thermal Properties of Ideal Gas}

\end{center}

\vskip .6cm
\medskip

\vspace*{4.0ex}

\baselineskip=18pt

\centerline{\large \rm  Rajesh Kumar Gupta, Meenu}

\vspace*{4.0ex}

\centerline{\large \it Department of Physics, Indian Institute of Technology Ropar, }

\centerline{\large \it  Rupnagar, Punjab 140001, India}

\vspace*{1.0ex}
\centerline{E-mail: rajesh.gupta@iitrpr.ac.in, meenu.20phz0003@iitrpr.ac.in }

\vspace*{5.0ex}

\centerline{\bf Abstract} \bigskip
In this article, we investigate the thermal properties of non-relativistic many-body systems at finite temperatures and chemical potential. We compute the one-point function of various operators constructed out of the basic fields in ideal bosonic and fermionic many-body systems. The one-point function is non-zero only for operators with zero particle numbers. We investigate these operators in $\mathbb R^d$ and $\mathbb R^d_{+}$, i.e. a flat space with a planar boundary. Furthermore, we compute the Green's function and using the operator product expansion, we express it in terms of the thermal one-point function of the higher spin currents. On $\mathbb R^d_{+}$, the operator product expansion allows to express the bulk-bulk Green's function in terms of the thermal Green's function of the boundary operators.

We also study the ideal system by placing it on curved spatial surfaces, specifically spherical surfaces.  We compute the partition function and Green's function on spheres, squashed-sphere and hemispheres. Finally, we compute the large radius corrections to the partition function and Green's function by expanding in the large radius limit.
\vfill \eject

\baselineskip=18pt

\tableofcontents
\section{Introduction}
Quantum field theory provides a framework to build models and compute various properties of quantum many-body systems. A non-relativistic many body system at a critical point has the scale symmetry. The quantum field theory description of such a system has the Galilean and anisotropic scaling symmetry. An interesting and special class of such a quantum field theory model is the unitary model which describes the system at unitarity. At the unitary limit, the system has enhanced symmetry, which is described by the Schr$\ddot{\text{o}}$dinger group. The Schr$\ddot{\text{o}}$dinger group is the group of space-time symmetry transformations consisting of an expansion transformation in addition to the scale and Galilean transformations, and a phase transformation whose generator, the particle number, appears as the central extension of the Galilean algebra~\cite{Hagen:1972pd, Henkel:1993sg, Mehen:1999nd, Nishida:2007pj}.

A natural thing to study is the phases of a non-relativistic many-body system at finite temperature and chemical potential. This can be investigated by studying non-relativistic field theory on S$^{1}_{\beta}\times\mathcal M^{d}$, where $\mathcal M^{d}$ is a spatial geometry.
One can computes observables such as thermal partition function and correlation functions as a function of temperature and chemical potentials. Moreover, various transport properties and response functions of the system at finite temperatures can be deduced by computing these observables on a non-trivial background. For example, the thermal partition function or the correlation function will be a function of the background fields and are generators of correlation functions with the insertion of currents that are dual to the background deformation. In this article, our goal will be to investigate various thermodynamic properties of a many-body system on a generic background. For this, we study non-relativistic field theories, describing Bosons or Fermions, on S$^{1}_{\beta}\times\mathcal M^{d}$.

Given a scalar operator $\mathcal O(\vec x)$ of particle number $m_{\mathcal O}$ and scaling dimension $\Delta_{\mathcal O}$, the space-time translational invariance and scale invariance implies that
\be\label{ThermalOnePt.1}
<\mathcal O(\tau,\vec x)>_{\beta}=b_{\mathcal O}\,\delta_{m_{\mathcal O},0}\,T^{\frac{\Delta_{\mathcal O}}{2}}\,.
\ee
Galilean symmetry implies that only scalar operators will have a non-zero one-point function, and furthermore, only the operators in the zero particle sector can have a non-zero thermal expectation value. 
In the presence of a dimensionful constant, say the chemical potential $\m$, the Galilean invariance and the scale invariance imply that
\be\label{ThermalOnePt.2}
<\mathcal O(\tau,\vec x)>_{\beta}=b_{\mathcal O}\,\delta_{m_{\mathcal O},0}\,T^{\frac{\Delta_{\mathcal O}}{2}}f(\frac{\m}{T^{\delta/2}})\,.
\ee
where $\delta$ being the dimension of the constant $\m$.
Our interests would be to compute the constant $b_{\mathcal O}$.

The zero particle sector is interesting in its own right. This sector includes operators like energy density, energy-momentum tensor and conserved currents of the theory. In a non-relativistic conformal field theory, the computation of vacuum correlation functions of these operators is subtle. This has to do with the fact that boost and translational generators in the sector commute with each other. As a result, the descendent of a primary operator with zero particle number, which is annihilated by the boost generator, can not be lowered back to the primary operator by the application of the boost generator. An important consequence of this observation is that we can not fix their correlation functions using the conformal Ward identities. Nevertheless, one can study the thermal one-point function of these operators. One of the motivations of the present work is to make quantitative statements about the thermal properties of these operators.

The operator product expansion (OPE) is a useful tool for studying the short-distance expansion of the product of operators. For example, such a short distance expansion has been useful to extract a new class of the universal relationship, called Tan-relations, among thermodynamic quantities in many-particle systems with short-range interactions~\cite{Braaten:2008uh, Braaten:2008bi, TAN20082952, TAN20082971}. In a non-relativistic field theory, where particle number is conserved, the OPE is constrained in the sense that the product of operators of particle number $m_{1}$ and $m_{2}$ has short distance expansion only in terms of operators of particle number $m_{1}+m_{2}$\cite{Golkar:2014mwa, Goldberger:2014hca, Gupta:2022azd},
\be
\mathcal O_{1}(\tau,\vec x)\times\mathcal O_{2}(0,\vec 0)\sim\sum_{\mathcal O}\sum_{r,s,q}f_{r,s,q}(\tau,\vec x)\mathcal O^{(r,s)}(0,\vec 0)\,,
\ee
where $\mathcal O^{(r,s,q)}(0,\vec 0)\sim x^{i_{1}}....x^{i_{r}}(i\p_{i_{1}})...(i\p_{i_{r}})(-\nabla^{2})^{s}(i\p_{\tau})^{q}\mathcal O(0,\vec 0)$ is an operator of particle number $m_{1}+m_{2}$.
Such OPE is useful in studying the correlation function at finite temperature. For example, the thermal one-point function of scalar operators can be extracted from the two-point function of scalar operators. A special case of the above OPE is the product of operators with the particle number $m_{1}=-m_{2}=m$. In this case, the expansion is in terms of operators with particle number zero. In this OPE, we expect all the currents to contribute. Focussing on non-interacting theory, one can obtain the OPE by simply doing the Taylor series expansion of the form:

\be\label{OPEZeroParticleSec}
\phi(\tau,\vec x)\times\phi^{\dagger}(0,\vec 0)\sim\sum_{\mathcal J^{(r,q)}_{s}}f_{r,s,q}(\tau,\vec x)\mathcal J^{(r,q)}_{i_{1}....i_{s}}x^{i_{1}}.....x^{i_{s}}\,. 
\ee
Here $f(\tau,\vec x)$ respects the symmetries of the system.
Since the left-hand side has a net zero particle number, the expansion on the right-hand side includes only those operators having zero particle numbers. These operators are constructed from the basic fields $\phi$ and $\phi^{\dagger}$. These are given by~\cite{Bekaert:2011qd}
\be
\mathcal J^{(r,q)}_{i_{1}....i_{s}}(0,\vec 0)=\phi^{\dagger}\underbrace{\overset{\leftrightarrow}{\partial_\tau}...\overset{\leftrightarrow}{\partial_\tau}}_{r}(\overset{\leftrightarrow}{\p}\cdot\overset{\leftrightarrow}{\p})^{q}\overset{\leftrightarrow}{\partial_{i_{1}}}....\overset{\leftrightarrow}{\partial_{i_{s}}}\phi(0,\vec 0)\,,
\ee
where 
\be
\phi^{\dagger}\overset{\leftrightarrow}\p_{\tau}\phi=\frac{1}{2}(\phi^{\dagger}\p_{\tau}\phi-\p_{\tau}\phi^{\dagger}\phi)\,.
\ee
Furthermore, to organize the expansion on the RHS of~\eqref{OPEZeroParticleSec}, we assume that the operators $\mathcal J^{(r,q)}_{i_{1}....i_{s}}(\tau,\vec x)$ are symmetric and traceless in its spin indices.
The thermal two-point function of the scalar operators, therefore, can be expressed in terms of the thermal one-point function of the operators $\mathcal J^{(r,q)}_{i_{1}....i_{s}}(\tau,\vec x)$. These one-point functions receive contributions from the trace part of the higher spin operators $\mathcal J^{(r,q)}_{i_{1}....i_{s}}(\tau,\vec x)$  which amounts to computing the one point function of the scalar operators of the form $\phi^{\dagger}\p_{\tau}^{r}(\p^{2})^{s+q}\phi$. Therefore, in the main texts, we compute the thermal expectation value of these operators. 

The thermal correlation function is sensitive to the deformations in the theory. For example, let's consider the system in the presence of a boundary. The thermal correlation function/one-point function will have the contributions from the degrees of freedom localized on the boundary. In other words, we expect the one-point function to have a similar form as~\eqref{ThermalOnePt.2}, except that the mathematical form for the function is much richer than the discussion without the boundary since the one-point function now receives a contribution from the boundary degrees of freedom.  To see this, suppose we consider a bCFT at finite temperature on $R_{+}^{d}$. We assume that the boundary is located at $y=0$ and the boundary coordinates are $\vec x$. Thus, the bulk space coordinates are $(\vec x,y)$. We have space and time translational invariance along the boundary direction. Thus, the translation invariance implies that the one point function of a scalar operator must be independent of $\tau$ and $\vec x$. Thus, we must have
\be
<\mathcal O(\tau, \vec x, y)>_{\beta}=\,\delta_{m_{\mathcal O},0}\,f_{\p}(y, \beta)\,.
\ee
Furthermore, the scale invariance implies that it should be of the form
\be
<\mathcal O(\tau,\vec x, y)>_{\beta}=\,\delta_{m_{\mathcal O},0}\,\frac{f_{\p}(Ty^{2})}{y^{\Delta}}=\,\delta_{m_{\mathcal O},0}\,T^{\frac{\Delta}{2}}\,\frac{f_{\p}(\xi)}{\xi^{\frac{\Delta}{2}}}\,.
\ee
Here $\xi=y^{2}T$. It will be interesting to obtain the form of $f_{\p}(\xi)$ in a non-relativistic conformal field theory in the presence of a boundary. In fact, a certain observation can be made about $f_{\p}(\xi)$ using the boundary OPE. In the presence of the planar boundary at $y=0$, we have the OPE~\cite{Gupta:2022azd}
\be
\mathcal O(\tau,\vec x, y)=\sum_{\hat{\mathcal O}}\frac{B^{\hat{\mathcal O}}_{\mathcal O}}{y^{\Delta_{\mathcal O}-\Delta_{\hat{\mathcal O}}}}D^{\Delta_{\hat{\mathcal O}}}(y^{2}\p^{2})\hat{\mathcal O}(\tau,\vec x)\,.
\ee
In the above, the notation $\p^{2}$ is the Schrodinger operator $\p^{2}=\p_{i}^{2}-2m_{\hat{\mathcal O}}\p_{\tau}$, the operator $D^{\Delta_{\hat{\mathcal O}}}(y^{2}\p^{2})$ represents the series of $y^{2}\p^{2}$ and the summation is over all boundary primary operators.
Thus, we have
\be
<\mathcal O(\vec x, y)>_{\beta}=\sum_{\hat{\mathcal O}}\frac{B^{\hat{\mathcal O}}_{\mathcal O}}{y^{\Delta_{\mathcal O}-\Delta_{\hat{\mathcal O}}}}D^{\Delta_{\hat{\mathcal O}}}(y^{2}\p^{2})<\hat{\mathcal O}(\vec x)>_{\beta}\,.
\ee
Now using the translational invariance in $\vec x$ and the scaling argument, we have
\be
<\hat{\mathcal O}(\tau,\vec x)>_{\beta}=\delta_{m_{\mathcal O},0}\,b_{\hat{\mathcal O}}T^{\frac{\Delta_{\hat{\mathcal O}}}{2}}\,,
\ee
and with the above, we obtain
\be
<\mathcal O(\vec x, y)>_{\beta}=\delta_{m_{\mathcal O},0}\,\sum_{\hat{\mathcal O}}\frac{B^{\hat{\mathcal O}}_{\mathcal O}}{y^{\Delta_{\mathcal O}-\Delta_{\hat{\mathcal O}}}}b_{\hat{\mathcal O}}T^{\frac{\Delta_{\hat{\mathcal O}}}{2}}=\delta_{m_{\mathcal O},0}\,\Big(\frac{T}{\xi}\Big)^{\frac{\Delta_{\mathcal O}}{2}}\sum_{\hat{\mathcal O}}B^{\hat{\mathcal O}}_{\mathcal O}b_{\hat{\mathcal O}}\xi^{\frac{\Delta_{\hat{\mathcal O}}}{2}}\,.
\ee
Thus, we have the relation
\be
f_{\p}(\xi)=\sum_{\hat{\mathcal O}}B^{\hat{\mathcal O}}_{\mathcal O}b_{\hat{\mathcal O}}\xi^{\frac{\Delta_{\hat{\mathcal O}}}{2}}\,
\ee
from which we can retrieve the expression for $B^{\hat{\mathcal O}}_{\mathcal O}b_{\hat{\mathcal O}}$.

Finally, we extend the above analysis, the computations of the thermal partition function and Green's function, on curved spatial surfaces such as $d$-dimensional sphere, $S^{d}$. The motivation for such a study is that, typically, we are interested in studying the properties of a system in a finite volume. It is important to understand the finite volume corrections to various physical quantities, including the effect on the phase transition. The sphere provides the simplest finite volume spatial geometry on which to study the field theory. We compute the thermal partition function and Green's function on two and three-dimensional spheres. In the infinite radius limit, we expect to obtain the result of that of the flat space~\footnote{See~\cite{David:2024pir} for a recent discussion in this direction for relativistic field theory. Some of the computations in the manuscript are influenced by calculations presented in that paper.}. However, the large radius expansion provides a systematic way to compute the finite volume corrections. 
At the end, we have also computed the partition function of the ideal Bose gas on a squashed S$^{3}$. The squashing is parametrized by the squashing parameter $\epsilon$ that encodes the deformation away from the S$^{3}$. The partition function, away from the S$^{3}$ partition function, has an expansion in terms of the thermal expectation value of the correlation function of the energy-momentum tensor.

The organization of the paper is as follow. In section~\ref{PartitionFn.FreeTheory}, we discuss the free bosonic and fermionic theory on flat space. Since the calculation of the section will be repeated in the later sections, we keep this section self contained. We compute the thermal one point function and  Green's function with and without a boundary. In section~\ref{FiniteTempSphere}, we repeated the calculation in the case of spheres. We compute the partition function on spheres in two and three dimensions and also computed the partition function on a three dimensional squashed sphere. Furthermore, we extend these calculations to hemispheres. In the last section~\ref{GreenFn}, we have computed the thermal Green's function on spheres and hemispheres. Finally, we conclude with a brief discussion in the section~\ref{Disc.}.
\section{Thermal one point function in free theories}\label{PartitionFn.FreeTheory}
In this section, we will compute the thermal one point function in the free field theory. 
We begin with the bosonic case and consider the following action
\be\label{Action0.1}
S=\int d\tau\,d^{d}x\,\phi^{\dagger}(\p_{\tau}-\frac{\nabla^{2}}{2m}+\m)\phi\,.
\ee
Here $\phi$ is a one component bosonic field and $\m$ is constant. We are interested in computing the partition function
\be
Z_{b}(\beta,\m)=e^{-\beta F_{b}(\beta,\m)}=\int[D\phi][D\phi^{\dagger}]\,e^{-S}\,.
\ee
In fact, from the partition function, we can obtain the thermal expectation value of number density
\be\label{OnePtFn.1}
<\phi^{\dagger}\phi>_{\beta,\m}=\frac{1}{V}\frac{\p F_{b}(\beta,\m)}{\p\m}\,,
\ee
and the energy density as
\be
<\mathcal H>_{\beta,z=e^{-\beta\m}}=-\frac{1}{V}\Big(F_{b}+\beta\frac{\p F_{b}}{\p \beta}\Big)\,.
\ee
Here $V$ is the volume of the $d$-dimensional space. These calculations are very straight forward. In the standard textbook, we have the free energy of an ideal bosonic system given by
\be
\beta F_{b}(\beta,\m)=V\int\frac{d^{d}k}{(2\pi)^{d}}\ln(1-e^{-\beta(\epsilon_{\vec k}+\m)})=-V\Big(\frac{m}{2\pi\beta}\Big)^{\frac{d}{2}}Li_{\frac{d}{2}+1}(e^{-\beta\m})\,.
\ee
From this, we obtain~\footnote{In this paper, we focused on the computation of the temperature dependent contributions to the one point functions. There could be contributions to the one point function from zero temperature part. }
\be
<\phi^{\dagger}\phi>_{\beta,\m}=\Big(\frac{m}{2\pi\beta}\Big)^{\frac{d}{2}}Li_{\frac{d}{2}}(e^{-\beta\m}),\quad <\mathcal H>_{\beta,z=e^{-\beta\m}}=\Big(\frac{m}{2\pi\beta}\Big)^{\frac{d}{2}}\frac{d}{2\beta}\,Li_{1+\frac{d}{2}}(e^{-\beta\m})\,.
\ee
Let us see how we arrive at the above expression of the free energy using the field theory calculation. We start by decomposing the field into Matsubara modes as 
\be
\phi(\tau,x)=T\sum_{n}\int_{\vec k}e^{-i\omega_{n}\tau}e^{i\vec k\cdot\vec x}\phi_{n}(\vec k)\,.
\ee
where $\omega_{n}=\frac{2\pi n}{\beta}$. Then the partition function is
\be
\ln Z_{b}(\beta,\m)=-V\sum_{n\in\mathbb Z}\int_{\vec k}\,\ln(-i\omega_{n}+\epsilon_{\vec k}+\m)\,,
\ee
where $\epsilon_{\vec k}=\frac{\vec k^{2}}{2m}$ and $\int_{\vec k}\equiv \int\frac{d^{d}k}{(2\pi)^{d}}$. Next, we sum over the Matsubara modes. To do this, we note that
\be
\frac{\p}{\p\m}\ln Z_{b}(\beta,\m)=-V\lim_{\varepsilon\rightarrow0+}\sum_{n\in\mathbb Z}\int_{\vec k}\,\frac{e^{i\omega_{n}\varepsilon}}{-i\omega_{n}+\epsilon_{\vec k}+\m}\,.
\ee
Here we have added the convergence factor $e^{i\omega_{n}\varepsilon}$. We can now sum over the Matsubara modes by introducing the function
\be
n_{B}(x)=\frac{1}{e^{i\beta x}-1}\,,
\ee
in which case, we have
\be
\frac{\p}{\p\m}\ln Z_{b}(\beta,\m)=-V\beta\lim_{\varepsilon\rightarrow0+}\oint\frac{dz}{2\pi}\int_{\vec k}\,\frac{e^{iz\varepsilon}}{-iz+\epsilon_{\vec k}+\m}n_{B}(z)
\ee
Here the contour is running around the real axis so as to pick up the poles at $z=\omega_{n}$. Deforming the contour, we can write
\bea
\frac{\p}{\p\m}\ln Z_{b}(\beta,\m)&=&-V\beta\lim_{\varepsilon\rightarrow0+}\int_{\vec k}\,\int_{-\infty-i0+}^{+\infty-i0+}\frac{dz}{2\pi}\frac{1}{e^{i\beta z}-1}\Big[\frac{e^{iz\varepsilon}}{-iz+\epsilon_{\vec k}+\m}+\frac{e^{-iz\varepsilon}}{iz+\epsilon_{\vec k}+\m}\Big]\nn\\
&&-V\beta\lim_{\varepsilon\rightarrow0+}\int_{\vec k}\,\int_{-\infty-i0+}^{+\infty-i0+}\frac{dz}{2\pi}\frac{e^{-iz\varepsilon}}{iz+\epsilon_{\vec k}+\m}\,.
\eea
Closing the contour in the lower half plane and picking up the pole at $z=-i(\epsilon_{\vec k}+\m)$, we obtain
\be
\frac{\p}{\p\m}\ln Z_{b}(\beta,\m)=-V\beta\int_{\vec k}\,\frac{1}{e^{\beta(\epsilon_{\vec k}+\m)}-1}\,.
\ee
Finally, integrating over $\m$, we get
\be
\ln Z_{b}(\beta,\m)=-V\int_{\vec k}\,\ln(1-e^{-\beta(\epsilon_{\vec k}+\m)})\,.
\ee
Note that in the above calculation we have assumed that $\m>0$. In the bosonic system, generically, there is no restriction for the chemical potential to satisfy $\m>0$. However, in either case, we get the same integral form for the partition function.

At this point, we can compute the thermal one point function of operators involving derivatives. To begin with, we start with the operators $\phi^{\dagger}\p_{\tau}\phi$ and $\phi^{\dagger}\nabla^{2}\phi$. In order to compute the thermal one point functions, we deform the original action by the term $\delta\alpha\,\phi^{\dagger}\p_{\tau}\phi$ and $\delta\g\,\phi^{\dagger}\nabla^{2}\phi$ and compute the partition function. We obtain
\be
\ln Z_{b}(\beta,\m;\delta\alpha,\delta\g)=\frac{V}{(2\pi)^{\frac{d}{2}}}\Big(\frac{(1+\delta\alpha)m}{\beta(1-\delta\g)}\Big)^{\frac{d}{2}}Li_{1+\frac{d}{2}}(e^{-\frac{\beta\m}{1+\delta\alpha}}).
\ee
From this expression, we obtain
\bea
&&<\phi^{\dagger}\p_{\tau}\phi>_{\beta,\m}=-\frac{1}{2\beta}\Big(\frac{m}{2\pi\beta}\Big)^{\frac{d}{2}}\Big(2\beta\m \,Li_{\frac{d}{2}}(e^{-\beta\m})+d\,Li_{1+\frac{d}{2}}(e^{-\beta\m})\Big)\,,\nn\\
&&<\phi^{\dagger}\nabla^{2}\phi>_{\beta,\m}=-\frac{d\,m}{\beta}\Big(\frac{m}{2\pi\beta}\Big)^{\frac{d}{2}}Li_{1+\frac{d}{2}}(e^{-\beta\m})\,.
\eea
It is easier to see that these one point functions satisfy
\be
<\phi^{\dagger}\p_{\tau}\phi>_{\beta,\m}-\frac{1}{2m}<\phi^{\dagger}\nabla^{2}\phi>_{\beta,\m}=-\m<\phi^{\dagger}\phi>_{\beta,\m}\,.
\ee
If we had performed the above calculations with anti-periodic boundary condition along the thermal circle, i.e. we are now considering fermions, then we would have obtained~\footnote{We note that the fermionic and bosonic partition functions do not cancel each other in the presence of the chemical potential. It would be interesting to see the effect of this in the context of supersymmetric non-relativistic field theory~\cite{Henkel:2005dj, Nakayama:2008qm, Nakayama:2008qz}.}
\be
\ln Z_{f}(\beta,\m;\delta\alpha,\delta\g)=-\frac{V}{(2\pi)^{\frac{d}{2}}}\Big(\frac{(1+\delta\alpha)m}{\beta(1-\delta\g)}\Big)^{\frac{d}{2}}Li_{1+\frac{d}{2}}(-e^{-\frac{\beta\m}{1+\delta\alpha}}).
\ee
From this expression, we obtain
\bea
&&<\phi^{\dagger}\p_{\tau}\phi>_{\beta,\m}=\frac{1}{2\beta}\Big(\frac{m}{2\pi\beta}\Big)^{\frac{d}{2}}\Big(2\beta\m \,Li_{\frac{d}{2}}(-e^{-\beta\m})+d\,Li_{1+\frac{d}{2}}(-e^{-\beta\m})\Big)\,,\nn\\
&&<\phi^{\dagger}\nabla^{2}\phi>_{\beta,\m}=\frac{d\,m}{\beta}\Big(\frac{m}{2\pi\beta}\Big)^{\frac{d}{2}}Li_{1+\frac{d}{2}}(-e^{-\beta\m})\,.
\eea
More generally, we can compute the thermal one point function of the operator $\phi^{\dagger}\p_{\tau}^{n}(\nabla^{2})^{p}\phi$. We follow the same strategy as before: we deform the action by the term $\delta\alpha\,\phi^{\dagger}\p_{\tau}^{n}(\nabla^{2})^{p}\phi$ and compute the partition function. We then differentiate the partition function with respect to $\delta\alpha$. In the case of periodic boundary condition, we obtain
\bea
<\phi^{\dagger}\p_{\tau}^{n}(-\frac{\nabla^{2}}{2m})^{p}\phi>_{\beta}=\frac{(-1)^{n}}{(2\pi)^{\frac{d}{2}}\Gamma(\frac{d}{2})}\frac{m^{\frac{d}{2}}}{\beta^{\frac{d}{2}+n+p}}\sum_{r=0}^{n}\frac{n!}{(n-r)!r!}(\m\beta)^{n-r}\Gamma(\frac{d}{2}+r+p)Li_{\frac{d}{2}+r+p}(e^{-\beta \m})\,.\nn\\
\eea
The result for the fermionic case trivially follows from the above expression.

One would wonder how does the above result depends on the critical exponent $z$. In the above discussion, we have considered the theory with the critical exponent 2, i.e $z=2$. To see the dependence of the thermal one point function on the critical exponent, we consider the following action
\be
S=\int d\tau\,d^{d}x\,\phi^{\dagger}(\p_{\tau}-\kappa(\nabla^{2})^{z}+\m)\phi\,.
\ee
The above action, in the absence of the chemical potential, has the scale invariance and the scale transformations on the coordinates are $\tau\rightarrow \lambda^{z}\tau$ and $\vec x\rightarrow \lambda\, \vec x$, where $\lambda\neq 0$. We repeat the same calculation as before and find that the one point function is (for the bosonic case)
\bea
<\phi^{\dagger}\p_{\tau}^{n}(-\nabla^{2})^{p}\phi>_{\beta}=\frac{(-1)^{n}}{(4\pi)^{\frac{d}{2}}\Gamma(\frac{d}{2})}\frac{1}{(\beta\kappa)^{\frac{d+2p}{2z}}\beta^{n}}\sum_{r=0}^{n}\frac{n!}{(n-r)!r!}(\beta\m)^{n-r}\Gamma(r+\frac{d+2p}{2z})Li_{r+\frac{d+2p}{2z}}(e^{-\beta\m})\,.\nn\\
\eea

We now come back to the original theory~\eqref{Action0.1} and compute the thermal two point function in the bosonic case.  The thermal two point function, following the operator product expansion, can be expressed in terms of the thermal one point function of higher spin operators. The two point function in the momentum space has the form
\be
G(i\omega_{n},\vec p)=-\frac{1}{i\omega_{n}-\epsilon_{p}-\m}\,.
\ee
In the position space, this becomes
\be\label{Flat-gExpr}
g(\tau,\vec x)=<T\phi(\tau,\vec x)\phi^{\dagger}(0,\vec 0)>_{\beta}=-T\sum_{n\in\mathbb Z}\int_{\vec p}\frac{e^{-i\omega_{n}\tau}e^{-i\vec p\cdot\vec x}}{i\omega_{n}-\epsilon_{p}-\m}\,.
\ee 
After performing the summation and integration over the momentum, we obtain
\be
g(\tau,\vec x)=\frac{1}{(2\pi)^{\frac{d}{2}}}\sum_{n\in\mathbb Z}\Theta(\tau+n\beta)\Big(\frac{m}{n\beta+\tau}\Big)^{\frac{d}{2}}e^{-\frac{mx^{2}+2\m\left(n\beta+\tau\right)^{2}}{2(n\beta+\tau)}}\,.
\ee
In order to arrive at the above expression, we have used the identity
\be
\sum_{n\in\mathbb Z}\delta(x-n)=\sum_{k\in\mathbb Z}e^{-2\pi ikx}\,,
\ee
and then convert the sum over $n$ in \eqref{Flat-gExpr} to summation over $k$. And then we perform the integration over $\vec p$.
Moreover, the above two point function is periodic 
\be
g(\tau,\vec x)=g(\tau-\beta,\vec x)\,.
\ee
Finally, expanding the temperature dependent term part of the thermal two point function in the powers of $x^{2}$ and $\tau$, for $0<\tau<\beta$, one finds that the expansion can be organized in the following form~\footnote{Note that there is a zero temperature contribution in the Green's function. However, our motivation is in the temperature dependent part.},
\be
g_{\text{temp.}}(\tau,\vec x)=\sum_{r=0}^{\infty}\sum_{n=0}^{\infty}\frac{(-1)^{r}}{r!n!}\frac{\Gamma(\frac{d}{2})}{\Gamma(\frac{d}{2}+r)}(x^{2})^{r}\tau^{n}\Big(\frac{m}{2}\Big)^{r}<\phi^{\dagger}\p_{\tau}^{n}(-\frac{\nabla^{2}}{2m})^{r}\phi>\,.
\ee
Finally, let us make comments about the Green's function in the presence of a $d-$dimensional spatial planar boundary. We consider $(d+1)$-dimensional bulk with $d$-dimensional planar boundary. The normal coordinate to the boundary is $y\geq 0$ and the boundary is located at $y=0$. Suppose, we consider the two point function of the bulk scalar operators in the non-interacting theory. Then, following the method of images, the bulk-bulk thermal Green's function with Dirichlet and Neumann boundary conditions are
\bea
g_{D,N}(\tau,\vec x,y_{1},y_{2})&=&<T\phi(\tau,\vec x,y_{1})\phi^{\dagger}(0,\vec 0,y_{2})>_{\beta}=\frac{1}{(2\pi)^{\frac{d}{2}}}\sum_{n\in\mathbb Z}\Theta(\tau+n\beta)\Big(\frac{m}{n\beta+\tau}\Big)^{\frac{d}{2}}e^{-\frac{mx^{2}+2\m\left(n\beta+\tau\right)^{2}}{2(n\beta+\tau)}}\times\nn\\
&&\qquad\qquad\qquad\qquad\qquad\qquad\qquad\qquad\qquad\times\Big(1\mp e^{-\frac{2my_{1}y_{2}}{n\beta+\tau}}\Big)\,.
\eea
Here $\text{`` $-$ '' (`` $+$ '')}$ sign inside the parenthesis is for Dirichlet (Neumann) boundary condition and $x^{2}=\vec x^{2}+(y_{1}-y_{2})^{2}$. 
Following the idea presented above, one can immediately obtain the thermal expectation value of bulk operators $<\phi^{\dagger}\p_{\tau}^{n}(-\frac{\nabla^{2}}{2m})^{r}\phi(y)>$ from the above expression of the Green's function. For example, the thermal expectation value of $<\phi^{\dagger}\phi(y)>$ and $<\phi^{\dagger}\p_{\tau}\phi(y)>$ are
\bea
&&<\phi^{\dagger}\phi(y)>_{D/N}=\Big(\frac{m}{2\pi\beta}\Big)^{\frac{d}{2}}\Big[Li_{\frac{d}{2}}(e^{-\beta\m})+\lambda \sum_{r=0}^{\infty}\frac{1}{r!}\Big(-\frac{2my^{2}}{\beta}\Big)^{r}Li_{\frac{d}{2}+r}(e^{-\beta\m})\Big]\,,\nn\\
&&<\phi^{\dagger}\p_{\tau}\phi(y)>_{D/N}=\Big(\frac{m}{2\pi\beta}\Big)^{\frac{d}{2}}\frac{1}{2\beta}\Big\{-\Big[d Li_{1+\frac{d}{2}}(e^{-\beta\m})+2\beta\m\, Li_{\frac{d}{2}}(e^{-\beta\m})\Big]\nn\\
&&\quad\quad\quad\quad+\lambda\sum_{r=0}^{\infty}\frac{(-1)^{r}}{r!}\Big(\frac{2my^{2}}{\beta}\Big)^{r}\Big[\frac{4my^{2}}{\beta}Li_{2+\frac{d}{2}+r}(e^{-\beta\m})-d Li_{1+\frac{d}{2}+r}(e^{-\beta\m})-2\beta\m Li_{r+\frac{d}{2}}(e^{-\beta\m})\Big]\Big\}\,.\nn\\
\eea
Here $\lambda=+1$ for Neumann and $-1$ for Dirichlet boundary conditions.

More generally, the thermal Green's function of the bulk scalar operators can be expressed in terms of the thermal Green's function of the boundary primary operators $\hat{\mathcal O}$. This we see by using the decomposition of bulk operators in terms of boundary primary operators as given by~\cite{Gupta:2022azd}
\be
\mathcal O(\tau,\vec x,y)=\sum_{\hat{\mathcal O}}\frac{B_{\mathcal O\hat{\mathcal O}}}{y^{\Delta_{\mathcal O}-\Delta_{\hat{\mathcal O}}}}\sum_{n=0}^{\infty}\frac{a_{n}}{n!}y^{2n}\mathcal D^{n}\hat{\mathcal O}(\tau,\vec x)\,,
\ee
where
\be
a_{n}=\frac{(-1)^{n}}{2^{2n}\Big(\Delta_{\hat{\mathcal O}}-\frac{d-1}{2}\Big)_{n}},\quad \mathcal D=\p_{i}^{2}-2m_{\hat{\mathcal O}}\p_{\tau}\,,
\ee
and $\hat{\mathcal O}$ are boundary primary operators. Then, the bulk-bulk Green's function is
\bea
G(\tau,\vec x,y)=<T\phi(\tau,\vec x,y_{1})\phi^{\dagger}(0,\vec 0,y_{2})>=\sum_{\hat{\mathcal O_{1}},\hat{\mathcal O}^{\dagger}_{2}}\frac{B_{\phi\hat{\mathcal O}}B_{\phi^{\dagger}\hat{\mathcal O}^{\dagger}_{2}}}{y_{1}^{\Delta_{\phi}-\Delta_{\hat{\mathcal O}_{1}}}y_{2}^{\Delta_{\phi^{\dagger}}-\Delta_{\hat{\mathcal O}^{\dagger}_{2}}}}\sum_{r,s=0}^{\infty}\frac{a_{r}a_{s}}{r!s!}y_{1}^{2r}y_{2}^{2s}\mathcal D_{1}^{r}\mathcal D_{2}^{s}\mathcal G_{\hat{\mathcal O}_{1}\hat{\mathcal O}^{\dagger}_{2}}(\tau,\vec x)\,.
\eea

\section{Finite temperature partition function on sphere and hemi-sphere}\label{FiniteTempSphere}
In this section, we will study the bosonic and fermionic ideal gas on a $d$-dimensional sphere at a finite temperature and chemical potential. More specifically, we will focus on non-relativistic field theories on two and three-dimensional spheres. As we emphasized in the introduction, the motivation is to compute the finite-size corrections to various physical quantities. We will compute the thermal partition function and Green's function in ideal bosonic and fermion theories. 
We consider the following action on a curved spatial surface
\be
S=\int_{0}^{\beta} d\tau\,\int d^{d}x\,\sqrt{g}\,\phi^{\dagger}\Big(\p_{\tau}-\frac{\nabla^{2}}{2}+\m\Big)\phi\,.
\ee
Here $g_{ij}$ is the metric on the spatial surface and $\nabla$ is the covariant derivative.
Note that there could be Ricci scalar term which we have absorbed in the definition of $\m$.
\subsection{Thermal partition function on $S^{2}$ and hemisphere}
We begin with the ideal Bosonic gas on a two-dimensional sphere, S$^{2}$, of radius $R$. We use the $(\theta,\phi)$-coordinates in which the metric of the sphere is
\be
ds^{2}=R^{2}(d\theta^{2}+\sin^{2}\theta\,d\phi^{2})\,.
\ee
Then we have the partition function 
\be
\ln Z_{b}=-\sum_{\ell=0}^{\infty}(2\ell+1)\ln\Big(1-e^{-\beta(\frac{(\ell+\frac{1}{2})^{2}}{r^{2}}+\tilde\m)}\Big)\,.
\ee
Here $r=\sqrt{2}R$ and $\tilde\m=\m-\frac{1}{4r^{2}}$.
It is very difficult to perform the above summation and obtain a closed-form expression. However, there is a nice trick, the Hubbard-Stratanovich trick, that one can use to perform the summation over $\ell$. Basically, in this trick, we linearize the quadratic term in the exponent which can then be summed over. Thus,
\bea
\ln Z_{b}&=&\sum_{\ell=0}^{\infty}\sum_{n=1}^{\infty}\frac{(2\ell+1)}{n}e^{-n\beta(\frac{(\ell+\frac{1}{2})^{2}}{r^{2}}+\tilde\m)}\,,\nn\\
&=&\frac{1}{2\sqrt{\pi \beta}}\sum_{\ell=0}^{\infty}\sum_{n=1}^{\infty}\frac{(2\ell+1)}{n^{\frac{3}{2}}}e^{-n\beta\tilde\m}\int_{-\infty-i\epsilon}^{\infty-i\epsilon}du\,e^{-\frac{u^{2}}{4n\beta}-\frac{i}{r}(\ell+\frac{1}{2})u}\,.
\eea
Now, we can perform the summation over $\ell$ and we obtain
\bea\label{S2partitionFn}
\ln Z_{b}&=&\frac{r}{2\sqrt{\pi \beta}}\sum_{n=1}^{\infty}\frac{1}{n^{\frac{3}{2}}}\int_{-\infty-i\epsilon}^{\infty-i\epsilon}du\,e^{-n\beta\tilde\m}\,e^{-\frac{u^{2}}{4n\beta}}\frac{\p}{\p u}\Big(\frac{1}{\sin(\frac{u}{2r})}\Big)\,,\nn\\
&=&\frac{r}{4\beta\sqrt{\pi \beta}}\sum_{n=1}^{\infty}\frac{1}{n^{\frac{5}{2}}}\int_{-\infty-i\epsilon}^{\infty-i\epsilon}du\,u\,e^{-n\beta\tilde\m}\,e^{-\frac{u^{2}}{4n\beta}}\frac{1}{\sin(\frac{u}{2r})}\,,
\eea
where in the last step, we have performed integrating by parts.
Now, instead of computing the integration explicitly, we obtain the large $r$-expansion of the above. Such an expansion we obtain by substituting the series expansion of sine function
\be
\frac{1}{\sin x}=\frac{1}{x}+\sum_{k=1}^{\infty}\frac{2(2^{2k-1}-1)|B_{2k}|x^{2k-1}}{(2k)!}\,,
\ee
into the integrand~\eqref{S2partitionFn} to obtain 
\bea\label{BosonicPartiS2}
\ln Z_{b}&=&\frac{r^{2}}{\beta}\sum_{n=1}^{\infty}\frac{1}{n^{2}}e^{-n\beta\tilde\m}\Big[1+\sum_{k=1}^{\infty}\frac{2(2^{2k-1}-1)}{(2r)^{2k}k!}|B_{2k}|(n\beta)^{k}\Big]\,,\nn\\
&=&\frac{r^{2}}{\beta}Li_{2}(e^{-\beta\tilde\m})+\frac{1}{12}Li_{1}(e^{-\beta\tilde\m})+\frac{7\beta}{480r^{2}}Li_{0}(e^{-\beta\tilde\m})+....
\eea
Note that the constant $\tilde\m$ could have dependence on $r$, in which case the polylogarithm also needs to be expanded in powers of $r^{-1}$. From the above expression for the partition function one can obtain various thermodynamic quantities in the large $r$-expansion. For example, the pressure and the energy density are
\bea
&&P=\frac{1}{2\pi \beta^{2}}\Big(Li_{2}(e^{-\beta\tilde\m})+\frac{\beta}{12r^{2}}Li_{1}(e^{-\beta\tilde\m})+\frac{7\beta^{2}}{480r^{4}}Li_{0}(e^{-\beta\tilde\m})+....\Big)\,,\nn\\
&&U=\frac{1}{2\pi \beta^{2}}\Big(Li_{2}(e^{-\beta\tilde\m})-\frac{7\beta^{2}}{480r^{4}}Li_{0}(e^{-\beta\tilde\m})+....\Big)\,.
\eea

The above analysis is easily extended to the case of anti-periodic boundary condition along the thermal circle, i.e. we look at the fermionic partition function. It is given as
\bea
\ln Z_{f}&=&\sum_{\ell=0}^{\infty}(2\ell+1)\ln\Big(1+e^{-\beta(\frac{(\ell+\frac{1}{2})^{2}}{r^{2}}+\tilde\m)}\Big)\,,\nn\\
&=&-\frac{r}{4\beta\sqrt{\pi \beta}}\sum_{n=1}^{\infty}\frac{(-1)^{n}}{n^{\frac{5}{2}}}\int_{-\infty-i\epsilon}^{\infty-i\epsilon}du\,u\,e^{-n\beta\tilde\m}\,e^{-\frac{u^{2}}{4n\beta}}\frac{1}{\sin(\frac{u}{2r})}\,.
\eea
So the final result we get
\bea\label{FermionicPartiS2}
\ln Z_{f}&=&-\frac{r^{2}}{\beta}Li_{2}(-e^{-\beta\tilde\m})-\frac{1}{12}Li_{1}(-e^{-\beta\tilde\m})-\frac{7\beta}{480r^{2}}Li_{0}(-e^{-\beta\tilde\m})-....
\eea
As a simple extension of the above results~\eqref{BosonicPartiS2} and~\eqref{FermionicPartiS2}, we can also study the theory on a hemisphere. Suppose we consider the hemisphere with the boundary located at $\theta=\frac{\pi}{2}$. In such case, we need to impose boundary conditions on the field. 
With the Dirichlet bounday condition, i.e. $\phi\Big|_{\theta=\frac{\pi}{2}}=0$, the partition function would be 
\be
\ln Z_{\p D}=-\sum_{\ell=0}^{\infty}\ell\ln\Big(1-e^{-\beta(\frac{(\ell+\frac{1}{2})^{2}}{r^{2}}+\tilde\m)}\Big)=\frac{1}{2}\ln Z_{S^{2}}+\frac{1}{2}\sum_{\ell=0}^{\infty}\ln\Big(1-e^{-\beta(\frac{(\ell+\frac{1}{2})^{2}}{r^{2}}+\tilde\m)}\Big)\,,
\ee
and with the Neumann boundary condition $\p_{y}\phi\Big|_{\theta=\frac{\pi}{2}}=0$, the partition function is
\be
\ln Z_{\p N}=-\sum_{\ell=0}^{\infty}(\ell+1)\ln\Big(1-e^{-\beta(\frac{(\ell+\frac{1}{2})^{2}}{r^{2}}+\tilde\m)}\Big)=\frac{1}{2}\ln Z_{S^{2}}-\frac{1}{2}\sum_{\ell=0}^{\infty}\ln\Big(1-e^{-\beta(\frac{(\ell+\frac{1}{2})^{2}}{r^{2}}+\tilde\m)}\Big)\,.
\ee
Here $Z_{S^{2}}$ is the partition function of the bosonic theory on S$^{2}$. One has the similar expression for the fermionic theory. Since we have already evaluated the first term, therefore, only the second term needed to be evaluated. Before that, we note that the partition function on the hemipshere is not exactly half of the sphere partition function. There is a correction to it due to the boundary which is torus in the present case. As we see below the second term in the partition function is exactly the partition function of a non-relativistic system with periodic boundary conditions on a torus.  We have
\bea
\sum_{\ell=0}^{\infty}\ln\Big(1-e^{-\beta(\frac{(\ell+\frac{1}{2})^{2}}{r^{2}}+\tilde\m)}\Big)&=&-\sum_{\ell=0}^{\infty}\sum_{n=1}^{\infty}\frac{1}{n}e^{-n\beta(\frac{(\ell+\frac{1}{2})^{2}}{r^{2}}+\tilde\m)}\,,\nn\\
&=&-\sum_{\ell=0}^{\infty}\sum_{n=1}^{\infty}\frac{1}{2\sqrt{\pi\beta}n^{\frac{3}{2}}}e^{-n\beta\tilde\m}\int_{-\infty-i\epsilon}^{\infty-i\epsilon}du\,e^{-\frac{u^{2}}{4n\beta}-i\frac{(\ell+\frac{1}{2})}{r}u}\,,\nn\\
&=&-\frac{1}{2\sqrt{\pi\beta}}\sum_{n=1}^{\infty}\frac{1}{n^{\frac{3}{2}}}e^{-n\beta\tilde\m}\int_{-\infty-i\epsilon}^{\infty-i\epsilon}du\,\frac{e^{-\frac{u^{2}}{4n\beta}}}{2i\sin\frac{u}{2r}}\,,\nn\\
&=&-\frac{1}{8i\sqrt{\pi\beta}}\sum_{n=1}^{\infty}\frac{1}{n^{\frac{3}{2}}}e^{-n\beta\tilde\m}\oint du\,\frac{e^{-\frac{u^{2}}{4n\beta}}}{\sin\frac{u}{2r}}\,,\nn\\
&=&-\frac{r}{2}\sqrt{\frac{\pi}{\beta}}\sum_{n=1}^{\infty}\frac{1}{n^{\frac{3}{2}}}e^{-n\beta\tilde\m}(1+2\sum_{p=1}^{\infty}(-1)^{p}e^{-\frac{p^{2}\pi^{2}r^{2}}{n\beta}})\,,\nn\\
&=&-\frac{r}{2}\sqrt{\frac{\pi}{\beta}}Li_{\frac{3}{2}}(e^{-\beta\tilde\m})-r\sqrt{\frac{\pi}{\beta}}\sum_{n=1}^{\infty}\frac{1}{n^{\frac{3}{2}}}e^{-n\beta\tilde\m}\sum_{p=1}^{\infty}(-1)^{p}e^{-\frac{p^{2}\pi^{2}r^{2}}{n\beta}}
\eea
We would like to mention a couple of comments here. Firstly, we observe that the final expression, the correction term due to the boundary, is precisely the same as the torus partition function of a non-relativistic system, see eq(4.14) in~\cite{Aguilera-Damia:2023jyc}.
Secondly, the boundary partition function contributes in the large $r$-corrections to the partition function; however, most of the terms are exponentially suppressed. Finally, we can compute the corrections to various thermodynamic quantities due to the presence of the boundary. For example, leading corrections to the pressure for the ideal bosonic system with the Dirichlet boundary conditions is
\be
P_{\p}=P-\frac{1}{4\pi r\sqrt{\pi\beta^{3}}}Li_{\frac{3}{2}}(e^{-\beta\tilde\m})\,.
\ee

\subsection{Thermal partition function on $S^{3}$ and 3-dimensional hemisphere}
Next, we consider the finite temperature partition function of the ideal Bose gas on $S^{3}$. The background metric has the following form
\be\label{3sphereMetric.1}
ds^{2}=R^{2}(d\chi^{2}+\sin^{2}\chi\,(d\theta^{2}+\sin^{2}\theta\,d\phi^{2}))\,,
\ee
where $\chi\in[0,\pi],\theta\in[0,\pi]$ and $\phi\in[0,2\pi)$. 
The eigen functions and eigen value of the Laplacian $\nabla_{S^{3}}^{2}$ on an unit $S^{3}$ are given by
\be
\nabla^{2}_{S^{3}}\Phi_{\ell,m,\sigma}=-\ell(\ell+2)\Phi_{\ell,m,\sigma}\,,
\ee
where $\ell=0,1,2,..,\infty$, $m=0,1,..,\ell$ and $\sigma=-m\, \text{to}\, +m$.  The explicit form of the eigen function is
\be
\Phi_{\ell,m,\sigma}(\chi,\theta,\phi)=\mathcal N_{\ell,m}\frac{1}{\sqrt{\sin\chi}}P_{\ell+\frac{1}{2}}^{-m-\frac{1}{2}}(\cos\chi)Y_{m,\sigma}(\theta,\phi)\,,
\ee
where the normalization factor is
\be
\mathcal N_{\ell,m}=\sqrt{\frac{(\ell-m)!}{(\ell+1)(\ell+m+1)!}}\,.
\ee
The total degeneracy for a given eigen value is
\be
d_{\ell}=(\ell+1)^{2}\,.
\ee
Thus, the thermal partition function is
\be\label{ThermPartitionS3.1}
\ln Z_{b}=-\sum_{\ell=0}^{\infty}d_{\ell}\ln(1-e^{-\beta\Big(\frac{\lambda_{\ell}}{r^{2}}+\m\Big)})=\sum_{\ell=0}^{\infty}(\ell+1)^{2}\sum_{n=1}^{\infty}\frac{1}{n}e^{-n\beta\Big(\frac{(\ell+1)^{2}}{r^{2}}+\tilde\m\Big)}\,,
\ee
where $\tilde\m=\m-\frac{1}{r^{2}}$ and $r=\sqrt{2}R$. We follow the same steps as we had done in the case of $S^{2}$. We obtain 
\bea
\ln Z_{b}&=&\sum_{\ell=0}^{\infty}(\ell+1)^{2}\sum_{n=1}^{\infty}\frac{1}{n}e^{-n\beta\tilde\m}\int_{-\infty-i\epsilon}^{\infty-i\epsilon}\frac{du}{2\sqrt{\pi n\beta}}e^{-\frac{u^{2}}{4n\beta}-\frac{i}{r}(\ell+1)u}\,,\nn\\
&=&-r^{2}\sum_{n=1}^{\infty}\frac{1}{n}e^{-n\beta\tilde\m}\int_{-\infty-i\epsilon}^{\infty-i\epsilon}\frac{du}{2\sqrt{\pi n\beta}}e^{-\frac{u^{2}}{4n\beta}}\frac{\p^{2}}{\p u^{2}}\frac{e^{-\frac{i}{2r}u}}{2i\sin\frac{u}{2r}}\,,\nn\\
&=&r^{2}\sum_{n=1}^{\infty}\frac{1}{32in^{\frac{7}{2}}\beta^{\frac{5}{2}}\sqrt{\pi}}e^{-n\beta\tilde\m}\oint\,du\,e^{-\frac{u^{2}}{4n\beta}}(2n\beta-u^{2})\frac{\cos(\frac{u}{2r})}{\sin\frac{u}{2r}}\,,\nn\\
&=&\frac{r^{3}\sqrt{\pi}}{4\beta^{\frac{5}{2}}}\sum_{n=1}^{\infty}\frac{1}{n^{\frac{7}{2}}}e^{-n\beta\tilde\m}\sum_{p\in\mathbb Z}e^{-\frac{r^{2}p^{2}\pi^{2}}{n\beta}}(n\beta-2r^{2}p^{2}\pi^{2})\,,\nn\\
&=&\frac{r^{3}\sqrt{\pi}}{4\beta^{\frac{3}{2}}}Li_{\frac{5}{2}}(e^{-\beta\tilde\m})+\frac{r^{3}\sqrt{\pi}}{2\beta^{\frac{5}{2}}}\sum_{n=1}^{\infty}\sum_{p=1}^{\infty}\frac{1}{n^{\frac{7}{2}}}e^{-n\beta\tilde\m-\frac{r^{2}p^{2}\pi^{2}}{n\beta}}(n\beta-2r^{2}p^{2}\pi^{2})\,.
\eea
In fact, the above can also be obtained directly from the Poisson resummation of~\eqref{ThermPartitionS3.1}~\footnote{The summation over the quantum number $\ell$ can be performed easily using the Poisson resummation. The sum is
\be
\sum_{\ell=1}^{\infty}\ell^{2}\,e^{-\frac{n\beta\ell^{2}}{r^{2}}}=-\frac{r^{2}}{2n}\frac{\p}{\p\beta}\sum_{\ell\in\mathbb Z}e^{-\frac{n\beta\ell^{2}}{r^{2}}}=-\frac{r^{3}}{2n}\frac{\p}{\p\beta}\Big(\sqrt{\frac{\pi}{n\beta}}\sum_{m\in\mathbb Z}e^{-\frac{\pi^{2}m^{2}r^{2}}{n\beta}}\Big)=\frac{r^{3}}{4}\frac{\sqrt{\pi}}{(n\beta)^{\frac{5}{2}}}\sum_{m\in\mathbb Z}e^{-\frac{\pi^{2}m^{2}r^{2}}{n\beta}}(n\beta-2m^{2}\pi^{2}r^{2})
\ee}.
The first term is the partition function on $\mathbb R^{3}$, and the rest of the terms are exponentially suppressed. This is in contrast to the 2-dimensional case where the finite $r$-correction to the partition function was polynomial in $\frac{1}{r}$, see for example~\eqref{BosonicPartiS2}. The extension to the fermionic case is straightforward and we will not repeat here.

One can extend the above analysis to the case of 3-dimensional hemisphere. We consider the boundary at $\chi=\frac{\pi}{2}$, which is a 2-dimensional sphere of radius $R$ with the induced metric is
\be\label{3sphereMetric.2}
ds^{2}=R^{2}\Big(d\theta^{2}+\sin^{2}\theta\,d\phi^{2}\Big)\,.
\ee
In the case, the partition function for the Dirichlet boundary condition is
\bea\label{ThermPartitionS3.2}
\ln Z_{\p D}&=&-\sum_{\ell=0}^{\infty}\frac{(\ell+2)(\ell+1)}{2}\sum_{n=1}^{\infty}\frac{1}{n}e^{-n\beta\Big(\frac{(\ell+1)^{2}}{r^{2}}+\tilde\m\Big)}\,,\nn\\
&=&\frac{1}{2}\ln Z_{S^{3}}-\frac{1}{2}\sum_{\ell=0}^{\infty}(\ell+1)\sum_{n=1}^{\infty}\frac{1}{n}e^{-n\beta\Big(\frac{(\ell+1)^{2}}{r^{2}}+\tilde\m\Big)}\,,
\eea
and for the Neumann's boundary condition is
\be
\ln Z_{\p N}=\frac{1}{2}\ln Z_{S^{3}}+\frac{1}{2}\sum_{\ell=0}^{\infty}(\ell+1)\sum_{n=1}^{\infty}\frac{1}{n}e^{-n\beta\Big(\frac{(\ell+1)^{2}}{r^{2}}+\tilde\m\Big)}\,.
\ee
Thus, we only need to evaluate the second term. Following the same strategy as we have done previously, we evaluate the second term in the large $r$-expansion. We obtain 
\be
\sum_{\ell=0}^{\infty}(\ell+1)\sum_{n=1}^{\infty}\frac{1}{n}e^{-n\beta\Big(\frac{(\ell+1)^{2}}{r^{2}}+\tilde\m\Big)}=\frac{r^{2}}{2\beta}\Big[Li_{2}(e^{\beta\tilde\m})-\sum_{n=1}^{\infty}\Big(\frac{\beta}{r^{2}}\Big)^{n}\frac{|B_{2n}|}{n!}Li_{2-n}(e^{-\beta\tilde\m})\Big]\,.
\ee
\subsection{Thermal partition function on squashed $S^{3}$}
Next, we would like to study the partition function of the ideal gas on a 3-dimensional squashed sphere. The motivation for considering the squashed 3-sphere is a two fold: the first is the computation of partition function on a less symmetric background. Squashing the sphere breaks the underlying spherical symmetry in a manner that preserve the rotation due to smaller set of the original symmetry group. For example, in the present case, we will be interested in the case of $S^{3}$, where the squashing breaks $SU(2)\times SU(2)\rightarrow SU(2)\times U(1)$ symmetry. The second is that the squashing provides a deformation parameter that can be used to probe the correlation function of energy-momentum current on S$^{3}$. 

We start with the metric of the squashed sphere having following form~\footnote{We are following the discussion presented in~\cite{Bobev:2017asb} where the case of relativistic CFTs was considered.}
\bea\label{Squ.Sphere}
ds^{2}&=&\frac{r^{2}}{4}(d\theta^{2}+\sin^{2}\theta\,d\phi^{2}+\frac{1}{1+\alpha}(d\eta+\cos\theta\,d\phi)^{2})\,,\nn\\
&=&\frac{r^{2}}{4}(d\theta^{2}+\sin^{2}\theta\,d\phi^{2}+(d\eta+\cos\theta\,d\phi)^{2})+\frac{\epsilon\, r^{2}}{4}(d\eta+\cos\theta\,d\phi)^{2}\,.
\eea
The parameter $\alpha$ is a constant with $\alpha>-1$ and the parameter $\epsilon=-\frac{\alpha}{1+\alpha}$ paramterizes the deformation away from the spherical geometry. We recover the unsquashed sphere $S^{3}$ for $\epsilon=0$. The Ricci scalar for the above metric is
\be
R=\frac{2}{r^{2}}(3-\epsilon)\,.
\ee
We, therefore, write the metric as
\be
g_{ij}=\bar g_{ij}+\epsilon\, h_{ij}\,\quad\text{and}\quad g^{ij}=\bar g^{ij}-\epsilon\, \kappa^{ij}\,,
\ee
where $k^{ij}=h^{ij}+\mathcal O(\epsilon)$ and $h^{ij}=\bar g^{ik}\bar g^{k\ell}h_{k\ell}$.
We want to compute the thermal partition function on the squashed sphere~\eqref{Squ.Sphere}. The free energy will be a function of the deformation parameter $\epsilon$. If we expand the partition function in powers of the deformation parameter then we have 
\bea\label{EpsilonExpansion}
Z^{\text{sq.}}(\beta,\epsilon)&=&Z_{0}(\beta)\Big[1+\frac{\epsilon}{2}\int_{0}^{\beta} d\tau\,\int\sqrt{\bar g}\,\kappa^{ij}(x)\Big<\Pi_{ij}(\tau,x)\Big>_{S^{3}}-\frac{\epsilon^{2}}{8}\int d\tau_{1}\int\sqrt{\bar g}\, \kappa(x)\kappa^{ij}(x)<\Pi_{ij}(\tau,x)>_{S^{3}}\nn\\
&&+\frac{\epsilon^{2}}{4}\int d\tau_{1}\,d\tau_{2}\int\sqrt{\bar g(x_{1})}\int\sqrt{\bar g(x_{2})} \kappa^{ij}(x_{1})\kappa^{kl}(x_{2})\Big<\frac{\delta \Pi_{ij}(\tau_{1},x_{1})}{\sqrt{\bar g(x_{2})}\delta g^{kl}(x_{2})}\Big>_{S^{3}}\nn\\
&&+\frac{\epsilon^{2}}{8}\int d\tau_{1}\,d\tau_{2}\int\sqrt{\bar g(x_{1})}\int\sqrt{\bar g(x_{2})}\kappa^{ij}(x_{1})\kappa^{kl}(\tau_{2},x_{2})\Big<\Pi_{ij}(\tau_{1},x_{1})\Pi_{kl}(\tau_{2},x_{2})\Big>_{S^{3}}....\Big]\,.\nn\\
\eea

Here, $<...>_{S^{3}}$ is the thermal expectation value on S$^{3}$ and $\bar g$ is the background metric, i.e. the metric on S$^{3}$. The energy-momentum current $\Pi_{ij}$ is defined by
\be
\Pi_{ij}=\frac{2}{\sqrt{g}}\frac{\delta S}{\delta g^{ij}}\,.
\ee
From the expression for the partition function~\eqref{EpsilonExpansion}, we see that the departure from the original S$^{3}$ partition function is encoded in the integral of the correlation functions of energy-momentum current of the theory.

For the squashed metric~\eqref{Squ.Sphere}, the non-zero component of $\kappa^{ij}$ is
\be
\kappa^{\eta\eta}=\frac{4}{r^{2}(1+\epsilon)}=\frac{4}{r^{2}}+\mathcal O(\epsilon)\,.
\ee
We will compute the partition function in ideal Bose gas. We consider the action to be
\be
S=\int dt\,d^{3}x\sqrt{g}\,\Big[\psi^{\dagger}\overset{\leftrightarrow}\p_{\tau}\psi+\frac{1}{2}g^{ij}\p_{i}\psi^{\dagger}\p_{j}\psi+\m\psi^{\dagger}\psi\Big]\,.
\ee
Then, we have
\be
\Pi_{ij}=\frac{1}{2}\Big(\p_{i}\psi^{\dagger}\p_{j}\psi+\p_{j}\psi^{\dagger}\p_{i}\psi\Big)-g_{ij}\Big(\psi^{\dagger}\overset{\leftrightarrow}\p_{\tau}\psi+\frac{1}{2}g^{kl}\p_{k}\psi^{\dagger}\p_{l}\psi+\m\psi^{\dagger}\psi\Big)\,.
\ee
Note that the above energy-momentum current is not traceless. 

The eigen value of the Laplacian $\nabla^{2}$ is given by
\be
\lambda_{\ell,q}=\ell(\ell+2)+\alpha(\ell-2q)^{2},\quad m_{\ell,q}=\ell+1\,,
\ee
where $0\leq q\leq \ell$. So, the thermal partition function is
\be
\ln Z^{\text{sq.}}_{b}=-\sum_{\ell,q}m_{\ell,q}\ln(1-e^{-\beta\Big(\frac{\lambda_{\ell,q}}{r^{2}}+\m\Big)})=\sum_{\ell=0}^{\infty}\sum_{q=0}^{\ell}(\ell+1)\sum_{n=1}^{\infty}\frac{1}{n}e^{-n\beta\Big(\frac{(\ell+1)^{2}+\alpha(\ell-2q)^{2}}{r^{2}}+\tilde\m\Big)}\,.
\ee
We first sum over $q$, we get
\be
\sum_{q=0}^{\ell}e^{-\frac{n\beta\alpha}{r^{2}}(\ell-2q)^{2}}=\int_{-\infty}^{\infty}\frac{du}{2\sqrt{\pi n\alpha\beta}}e^{-\frac{u^{2}}{4n\beta\alpha}}\frac{1}{2i\sin\frac{u}{r}}(e^{i\frac{u}{r}(1+\ell)}-e^{-\frac{iu}{r}(\ell+1)})\,.
\ee
Next, we sum over $\ell$. Following the same steps as done previously, we obtain the integral representation for the squashed partition function. It is
\be
\ln Z^{\text{sq.}}_{b}=ir\sum_{n=1}^{\infty}\frac{1}{16n^{3}\pi\beta^{2}\sqrt{\alpha}}e^{-n\beta\tilde\m}\int_{-\infty}^{\infty}du\,e^{-\frac{u^{2}}{4n\beta\alpha}}\int_{-\infty-i\epsilon}^{\infty-i\epsilon}dv\,v\,e^{-\frac{v^{2}}{4n\beta}}\frac{1}{\cos\frac{v}{r}-\cos\frac{u}{r}}\,.
\ee
Here we have two integrations: one due to the sum over $q$ and another due to the sum over $\ell$.
Now, we evaluate the above integral in the large $r$-expansion. In order to do that, we consider the following large $r$-expansion, 
\be\label{uv-large-rexpansion}
\frac{1}{\cos\frac{v}{r}-\cos\frac{u}{r}}=\frac{2r^{2}}{u^{2}-v^{2}}+\frac{u^{2}+v^{2}}{6(u^{2}-v^{2})}+\frac{3u^{4}+8u^{2}v^{2}+3v^{4}}{360r^{2}(u^{2}-v^{2})}+....=\frac{f(u,v)}{u^{2}-v^{2}}\,,
\ee
where $f(u,v)$ is smooth and even function in both $u$ and $v$ and is given by
\be
f(u,v)=2r^{2}+\frac{u^{2}+v^{2}}{6}+\frac{3u^{4}+8u^{2}v^{2}+3v^{4}}{360r^{2}}+.....\,.
\ee
An important point worth to mention here. Note that the way we have done the large $r$-expansion in~\eqref{uv-large-rexpansion}, the RHS has poles only at $v=\pm u$. Clearly, the denominator on the LHS of~\eqref{uv-large-rexpansion} has more poles, i.e.  at $v=\pm u+ 2\pi n r$ for some $n\in \mathbb Z$. It is not difficult to see that only the poles $v=\pm u$ give rise $\frac{1}{r}$-expansion. Other poles give exponentially suppressed corrections~\footnote{Suppose we include the pole $v=\pm u+2\pi r$. Then after performing the $v$-integration, we have the terms $\propto e^{-\frac{\pi^{2}r^{2}}{n\beta}}$. This is exponentially suppressed in the large $r$-limit.}. 

Next, we evaluate the $v$-integral. Using the integration
\bea
\int_{-\infty-i\epsilon}^{\infty-i\epsilon}dv\,v\,e^{-\frac{v^{2}}{4n\beta}}\frac{f(u,v)}{u^{2}-v^{2}}
=-\pi ie^{-\frac{u^{2}}{4n\beta}}\tilde f(u)\,,
\eea
where
\be
\tilde f(u)=f(u,v=u)=2r^{2}+\frac{u^{2}}{3}+\frac{7u^{4}}{180r^{2}}+....
\ee
we have the partition function
\bea
\ln Z^{\text{sq.}}_{b}&=& r\sum_{n=1}^{\infty}\frac{1}{16n^{3}\beta^{2}\sqrt{\alpha}}e^{-n\beta\tilde\m}\int_{-\infty}^{\infty}du\,e^{-\frac{u^{2}}{4n\beta\alpha}}e^{-\frac{u^{2}}{4n\beta}}\tilde f(u)=r\sum_{n=1}^{\infty}\frac{1}{16n^{3}\beta^{2}\sqrt{\alpha}}e^{-n\beta\tilde\m}\int_{-\infty}^{\infty}du\,e^{\frac{u^{2}}{4n\beta\epsilon}}\tilde f(u)\,,\nn\\
&=&\frac{r^{3}\sqrt{\pi}}{4\beta^{\frac{3}{2}}}\Big[Li_{\frac{5}{2}}(e^{-\beta\tilde\m})+\frac{\epsilon}{6r^{2}}(-3r^{2}Li_{\frac{5}{2}}(e^{-\beta\tilde\m})+2\beta Li_{\frac{3}{2}}(e^{-\beta\tilde\m}))+\mathcal O(\epsilon^{2})\Big]\,.
\eea

\section{Green's function on sphere}\label{GreenFn}
In this section, we will compute Green's function on spheres in two and three dimensions. We will also compute the Green's function on hemispheres. We will study these in the large $r$-expansion.
\subsection{Green's function on $S^{1}\times S^{2}$}
We will start with the Green's function on two dimensional sphere of radius $R$.
We want to solve the equation
\be
(\p_{\tau}-\frac{\nabla^{2}}{2}+\tilde\m)G(\tau;x,y)=\delta(\tau)\delta^{2}(x-y)\,.
\ee
The normalized eigen function of the differential operator is $\frac{1}{r\sqrt{\beta}}e^{-i\omega_{n}\tau}Y_{\ell}^{m}(\theta,\phi)$ with eigen value $(-i\omega_{n}+\frac{\ell(\ell+1)}{r^{2}}+\tilde\m)$, where $\omega_{n}=\frac{2\pi n}{\beta}$. 
Here $Y_{\ell}^{m}(\theta,\phi)$ satisfies the orthogonality property
\be
\int_{0}^{\pi}d\theta\,\sin\theta\,\int_{0}^{2\pi}d\phi\,Y_{\ell}^{m\,*}(\theta,\phi)Y_{\ell'}^{m'}(\theta,\phi)=\delta_{\ell,\ell'}\delta_{m,m'}\,.
\ee
Then, the Green's function is
\be
G(\tau;\theta_{1},\phi_{1},\theta_{2},\phi_{2})=\frac{1}{r^{2}\beta}\sum_{n\in\mathbb Z}\sum_{\ell=0}^{\infty}\sum_{m=-\ell}^{m=\ell}\frac{e^{-i\omega_{n}\tau}}{-i\omega_{n}+\frac{\ell(\ell+1)}{r^{2}}+\m}Y_{l}^{m\,*}(\theta_{1},\phi_{1})Y_{\ell}^{m}(\theta_{2},\phi_{2})\,.
\ee
Using the addition theorem of the spherical harmonics
\be
P_{\ell}(\cos\g)=\frac{4\pi}{2\ell+1}\sum_{m=-\ell}^{m=\ell}Y_{l}^{m\dagger}(\theta_{1},\phi_{1})Y_{\ell}^{m}(\theta_{2},\phi_{2})\,,
\ee
where $\g$ is the angle between two unit vectors and is given by
\be
\cos\g=\cos\theta_{1}\,\cos\theta_{2}+\sin\theta_{1}\,\sin\theta_{2}\,\cos(\phi_{1}-\phi_{2})\,,
\ee
we have the expression for the Green's function given as
\be
G(\tau;\theta_{1},\phi_{1},\theta_{2},\phi_{2})=\frac{1}{4\pi r^{2}\beta}\sum_{n\in\mathbb Z}\sum_{\ell=0}^{\infty}(2\ell+1)\frac{e^{-i\omega_{n}\tau}}{-i\omega_{n}+\frac{\ell(\ell+1)}{r^{2}}+\m}P_{\ell}(\cos\g)\,.
\ee
Now, we perform the sum over Matsubara frequency. Using the identity
\be
\sum_{k\in\mathbb Z}e^{-2\pi ik x}=\sum_{n\in\mathbb Z}\delta(x-n)\,,
\ee
we can write
\bea
\sum_{n\in\mathbb Z}\frac{e^{-i\omega_{n}\tau}}{-i\omega_{n}+\frac{\ell(\ell+1)}{r^{2}}+\m}&=&\sum_{n\in\mathbb Z}\int_{-\infty}^{\infty} d\omega\,\delta(\omega-n)\,\frac{e^{-\frac{2\pi i}{\beta}\omega\tau}}{-\frac{2\pi i}{\beta}\omega+\frac{\ell(\ell+1)}{r^{2}}+\m}\,,\nn\\
&=&\sum_{k\in\mathbb Z}\int_{-\infty}^{\infty} d\omega\,e^{-2\pi ik\omega}\,\frac{e^{-\frac{2\pi i}{\beta}\omega\tau}}{-\frac{2\pi i}{\beta}\omega+\frac{\ell(\ell+1)}{r^{2}}+\m}\,,\nn\\
&=&\sum_{k\in\mathbb Z}g(\tau+k\beta)\,,
\eea
where
\be
g(\tau)=\int_{-\infty}^{\infty}\,d\omega\,\frac{e^{-\frac{2\pi i}{\beta}\omega\tau}}{-\frac{2\pi i}{\beta}\omega+\frac{\ell(\ell+1)}{r^{2}}+\m}\,.
\ee
Thus, we can write the Green's function as the sum over images 
\be
G(\tau;\theta_{1},\phi_{1},\theta_{2},\phi_{2})=\frac{1}{4\pi r^{2}\beta}\sum_{k\in\mathbb Z}g(\tau+k\beta;\g)\,,
\ee
where
\be
g(\tau;\g)=\sum_{\ell=0}^{\infty}(2\ell+1)\int_{-\infty}^{\infty} d\omega\,\frac{e^{-\frac{2\pi i}{\beta}\omega\tau}}{-\frac{2\pi i}{\beta}\omega+\frac{\ell(\ell+1)}{r^{2}}+\m}P_{\ell}(\cos\g)\,.
\ee
Performing the integration over $\omega$ (for $\m>0$), we obtain
\bea
g(\tau;\g)&=&\Theta(\tau)\beta\sum_{\ell=0}^{\infty}(2\ell+1)e^{-\tau(\frac{\ell(\ell+1)}{r^{2}}+\m)}P_{\ell}(\cos\g)\,,\nn\\
&=&\Theta(\tau)\beta\sum_{\ell=0}^{\infty}(2\ell+1)e^{-\tau\tilde\m}P_{\ell}(\cos\g)\int_{-\infty+i\epsilon}^{\infty+i\epsilon}\frac{du}{2\sqrt{\pi\tau}}e^{-\frac{u^{2}}{4\tau}+i\frac{\ell+\frac{1}{2}}{r}u}\,.
\eea
Now, the sum over $\ell$ can be performed as follows:
\be
\sum_{\ell=0}^{\infty}(2\ell+1)P_{\ell}(\cos\g)e^{+i\frac{\ell+\frac{1}{2}}{r}u}=-\sqrt{2}\,ir\frac{\p}{\p u}\frac{1}{\sqrt{\cos\frac{u}{r}-\cos\g}}\,.
\ee
Thus, after performing the integration by parts, we can write
\bea\label{Green'sFng-S2.1}
g(\tau;\g)&=&-\Theta(\tau)\frac{ir\beta}{\sqrt{2\pi\tau}}e^{-\tau\tilde\m}\int_{-\infty+i\epsilon}^{\infty+i\epsilon}du\,\frac{u}{2\tau}\,\frac{e^{-\frac{u^{2}}{4\tau}}}{\sqrt{\cos\frac{u}{r}-\cos\g}}\,,\nn\\
&=&-\Theta(\tau)\frac{ir\beta}{4\sqrt{2\pi}\tau^{\frac{3}{2}}}e^{-\tau\tilde\m}\oint du\,u\,\frac{e^{-\frac{u^{2}}{4\tau}}}{\sqrt{\cos\frac{u}{r}-\cos\g}}\,.
\eea
In the above, we have the contour running in the clockwise direction around the real $u$-axis. The integrand has a branch cut, however, we will not evaluate the integral and obtain the result in a closed form expression. We will perform the integration in the large $r$ limit. It turned out that if we perform the integration keeping $\g$ fixed while taking $r\rightarrow\infty$ limit, the result vanishes. On the other hand, we obtain a non-trivial answer when $\tilde\g=r\g$ is kept fixed while taking $r\rightarrow\infty$. In this limit, we use the following series expansion
\be
\frac{1}{\sqrt{\cos\frac{u}{r}-\cos\g}}=\frac{\sqrt{2}r}{\sqrt{\tilde\g^{2}-u^{2}}}\Big(1+\frac{u^{2}+\tilde\g^{2}}{24r^{2}}+\frac{7u^{4}+22u^{2}\tilde\g^{2}+7\tilde\g^{4}}{5760 r^{4}}+....\Big)\,.
\ee
Plugging this in~\eqref{Green'sFng-S2.1}, we have
\be\label{Green'sFng-S2.2}
g(\tau;\g)=-\Theta(\tau)\frac{ir^{2}\beta}{4\sqrt{\pi}\tau^{\frac{3}{2}}}e^{-\tau\tilde\m}\oint du\,u\,\frac{e^{-\frac{u^{2}}{4\tau}}}{\sqrt{\tilde\g^{2}-u^{2}}}f(u,\tilde\g,r)\,,
\ee
where
\be
f(u,\tilde\g,r)=1+\frac{u^{2}+\tilde\g^{2}}{24r^{2}}+\frac{7u^{4}+22u^{2}\tilde\g^{2}+7\tilde\g^{4}}{5760 r^{4}}+....\,.
\ee
The integral~\eqref{Green'sFng-S2.2} has branch cut at $u=\pm\tilde\g$. Choosing the branches $(-\infty, -\tilde\g)$ and $(+\tilde\g,\infty)$ and simplifying the integral, we obtain
\bea
g(\tau;\g)&=&\Theta(\tau)\frac{r^{2}\beta}{\sqrt{\pi}\tau^{\frac{3}{2}}}e^{-\tau\tilde\m}\int_{0}^{\infty}dx\,(x+\tilde\g)\frac{e^{-\frac{(x+\tilde\g)^{2}}{4\tau}}}{\sqrt{x(2\tilde\g+x)}}f(u=x+\tilde\g,\tilde g,r)\,,\nn\\
&=&\Theta(\tau)\frac{r^{2}\beta}{\sqrt{\pi}\tau^{\frac{3}{2}}}e^{-\tau\tilde\m}\tilde\g\int_{1}^{\infty}dx\,x\frac{e^{-\frac{\tilde\g^{2}x^{2}}{4\tau}}}{\sqrt{(x+1)(x-1)}}\Big(1+\frac{\tilde\g^{2}(x^{2}+1)}{24r^{2}}+....\Big)\,,\nn\\
&=&\Theta(\tau)\frac{r^{2}\beta}{\tau}e^{-\tau\tilde\m}e^{-\frac{\tilde\g^{2}}{4\tau}}\Big[1+\frac{\tilde\g^{2}+\tau}{12r^{2}}+...\Big]\,.
\eea
Thus, the Green's function in the large $r$-expansion is 
\bea
G(\tau;\theta_{1},\phi_{1},\theta_{2},\phi_{2})&=&\frac{1}{4\pi r^{2}\beta}\sum_{k\in\mathbb Z}g(\tau+k\beta;\g)\,,\nn\\
&=&\frac{1}{4\pi }\sum_{k\in\mathbb Z}\Theta(\tau+k\beta)\frac{e^{-\frac{\tilde\g^{2}+4(\tau+k\beta)^{2}\tilde\m}{4(\tau+k\beta)}}}{(\tau+k\beta)}\Big[1+\frac{\tilde\g^{2}+(\tau+k\beta)}{12r^{2}}+...\Big]\,.\nn\\
\eea
Note that, the first term in the large $r$-expansion of the Green's function is the Green's function on the flat space, $\mathbb R^{2}$. The second and dot terms give the large $r$-correction to the Green's function.

Finally, we can also write down the expression for the Green's function on hemisphere. Basically, following the method of images, the Dirichlet/Neumann Green's function will be
\be
G_{D/N}(\tau;\theta_{1},\phi_{1},\theta_{2},\phi_{2})=G(\tau;\theta_{1},\phi_{1},\theta_{2},\phi_{2})\mp G(\tau;\theta_{1},\phi_{1},\pi-\theta_{2},\phi_{2}),.
\ee
Following the above analysis, one can also very easily perform the large $r$-expansion for the hemisphere Green's function. 
\subsection{Green's function on $S^{1}\times S^{3}$}
Next, we continue the computation of the thermal Green's function on the sphere $S^{3}$ of the radius $R$. The normalized eigen functions of the Schrodinger operator is $\frac{1}{\sqrt{r^{3}\beta}}e^{-i\omega_{n}\tau}\Phi_{L,\ell,m}(\theta,\phi)$ with eigen value $(-i\omega_{n}+\frac{\ell(\ell+2)}{r^{2}}+\tilde\m)$, where $\omega_{n}=\frac{2\pi n}{\beta}$. 
Thus, the Green's function is
\be
G(\tau;\chi_{1},\theta_{1},\phi_{1},\chi_{2},\theta_{2},\phi_{2})=\frac{1}{r^{3}\beta}\sum_{n\in\mathbb Z}\sum_{\ell=0}^{\infty}\sum_{m=0}^{\ell}\sum_{\sigma=-m}^{m}\frac{e^{-i\omega_{n}\tau}}{-i\omega_{n}+\frac{\ell(\ell+2)}{r^{2}}+\m}\Phi_{l,m,\sigma}^{*}(\chi_{1},\theta_{1},\phi_{1})\Phi_{\ell,m,\sigma}(\chi_{2},\theta_{2},\phi_{2})
\ee
Now, using the addition theorem of spherical harmonics, we obtain~\cite{Frye:2012jj}
\be
\sum_{m=0}^{\ell}\sum_{\sigma=-m}^{m}\Phi_{l,m,\sigma}^{*}(\chi_{1},\theta_{1},\phi_{1})\Phi_{\ell,m,\sigma}(\chi_{2},\theta_{2},\phi_{2})=\frac{(\ell+1)^{2}}{2\pi^{2}}\tilde P_{\ell}(\cos\g)\,,
\ee
where $\g$ is the angle between two vectors on $S^{3}$ whose explicit form is
\be
\g=\cos\chi_{1}\,\cos\chi_{2}+\sin\chi_{1}\,\sin\chi_{2}\,\Big(\cos\theta_{1}\,\cos\theta_{2}+\sin\theta_{1}\,\sin\theta_{2}\,\cos(\phi_{1}-\phi_{2})\Big)
\ee
 Here $\tilde P_{\ell}(\cos\g)$ is the Legendre polynomial satisfying the differential equation
\be
(1-x^{2})\frac{d^{2}}{dx^{2}}\tilde P_{\ell}(x)-3x\,\frac{d}{dx}\tilde P_{\ell}(x)+\ell(\ell+2)\tilde P_{\ell}(x)=0\,.
\ee
Then the Green's function is
\be
G(\tau;\chi_{1},\theta_{1},\phi_{1},\chi_{2},\theta_{2},\phi_{2})=\frac{1}{2\pi^{2}r^{3}\beta}\sum_{n\in\mathbb Z}\sum_{\ell=0}^{\infty}(\ell+1)^{2}\frac{e^{-i\omega_{n}\tau}}{-i\omega_{n}+\frac{\ell(\ell+2)}{r^{2}}+\m}\tilde P_{\ell}(\cos\g)
\ee
Now, we can repeat the same process as before.
Thus, we can write the Green's function as
\be
G(\tau;\chi_{1},\theta_{1},\phi_{1},\chi_{2},\theta_{2},\phi_{2})=\frac{1}{2\pi^{2} r^{3}\beta}\sum_{k\in\mathbb Z}g(\tau+k\beta;\g)\,,
\ee
where
\be
g(\tau;\g)=\sum_{\ell=0}^{\infty}(\ell+1)^{2}\int_{-\infty}^{\infty} d\omega\,\frac{e^{-\frac{2\pi i}{\beta}\omega\tau}}{-\frac{2\pi i}{\beta}\omega+\frac{\ell(\ell+2)}{r^{2}}+\m}\tilde P_{\ell}(\cos\g)\,.
\ee
Performing the integration over $\omega$, we obtain
\bea
g(\tau;\g)&=&\Theta(\tau)\beta\sum_{\ell=0}^{\infty}(\ell+1)^{2}e^{-\tau(\frac{\ell(\ell+2)}{r^{2}}+\m)}\tilde P_{\ell}(\cos\g)\,,\nn\\
&=&\beta\Theta(\tau)\sum_{\ell=0}^{\infty}(\ell+1)^{2}e^{-\tau\tilde\m}\tilde P_{\ell}(\cos\g)\int_{-\infty+i\epsilon}^{\infty+i\epsilon}\frac{du}{2\sqrt{\pi\tau}}e^{-\frac{u^{2}}{4\tau}+i\frac{\ell+1}{r}u}\,.
\eea
Now, in order to sum over $\ell$, we using the following useful relation
\be
\sum_{\ell=0}^{\infty}(\ell+1)^{2}x^{\ell}\tilde P_{\ell}(t)=\frac{1-x^{2}}{(1-2xt+x^{2})^{2}}\,.
\ee
Now, using the above relation, we obtain
\be
\sum_{\ell=0}^{\infty}(\ell+1)^{2}\tilde P_{\ell}(\cos\g)e^{i\frac{\ell+1}{r}u}=-\frac{i\sin\frac{u}{r}}{2(\cos\frac{u}{r}-\cos\g)^{2}}\,.
\ee
Thus, we get
\be
g(\tau;\g)=-i\beta r\,\Theta(\tau)\,e^{-\tau\tilde\m}\int^{\infty+i\epsilon}_{-\infty+i\epsilon}\frac{du}{2\sqrt{\pi\tau}}\frac{u}{4\tau}\frac{e^{-\frac{u^{2}}{4\tau}}}{\cos\frac{u}{r}-\cos\g}\,.
\ee
Next, we perform the integration over $u$. The above integration can be written as a contour integration as
\be
g(\tau;\g)=\frac{i\beta r}{16\tau\sqrt{\pi\tau}}\,\Theta(\tau)\,e^{-\tau\tilde\m}\oint\,du\,u\frac{e^{-\frac{u^{2}}{4\tau}}}{\cos\frac{u}{r}-\cos\g}\,,
\ee
where the contour encloses the real $u$-axis in the anti-clockwise direction. We have poles at $u=\pm r\g\pm 2rn\pi$. Clearly, poles for $n\neq 0$ give exponentially suppressed corrections to the Green's function. Furthermore, if we take $r\rightarrow\infty$, in which case, we keep $\tilde\g=r\g$ fixed, then, we get
\bea
g(\tau;\g)&=&\Theta(\tau)\frac{\beta\sqrt{\pi} }{4\tau^{\frac{3}{2}}}\,\frac{r^{3}\g}{\sin\g}e^{-\tau\tilde\m}e^{-\frac{r^{2}\g^{2}}{4\tau}}+\text{exp. suppr. terms}\nn\\
&=&\Theta(\tau)\frac{\beta\sqrt{\pi} }{4\tau^{\frac{3}{2}}}\,\frac{r^{2}\tilde\g}{\sin\g}e^{-\tau\tilde\m}e^{-\frac{\tilde\g^{2}}{4\tau}}+\text{exp. suppr. terms}\nn\\
&=&\Theta(\tau)\frac{\beta\sqrt{\pi} }{4\tau^{\frac{3}{2}}}\,e^{-\tau\tilde\m}e^{-\frac{\tilde\g^{2}}{4\tau}}r^{3}\Big(1+\sum_{k=1}^{\infty}\frac{2(2^{2k-1}-1)|B_{2k}|}{(2k)!}\Big(\frac{\tilde\g}{r}\Big)^{2k}\Big)+\text{exp. suppr. terms}
\eea
Finally, we can write down the expression for the Green's function on three dimensional hemisphere. Basically, following the method of images, the Dirichlet/Neumann Green's function will be
\be
G(\tau;\chi_{1},\theta_{1},\phi_{1},\chi_{2},\theta_{2},\phi_{2})_{D/N}=G(\tau;\chi_{1},\theta_{1},\phi_{1},\chi_{2},\theta_{2},\phi_{2})\mp G(\tau;\chi_{1},\theta_{1},\phi_{1},\pi-\chi_{2},\theta_{2},\phi_{2}),.
\ee
Following the previous analysis, one can also very easily perform the large $r$-expansion for the above hemisphere Green's function. 

\section{Discussion}\label{Disc.}
In this article, we have studied the thermal properties of non-interacting, non-relativistic systems. Our motivation is to explore the thermodynamics of the many-body system on a generic background. Here, we have focussed only on the free theories. We have computed the partition function and one-point function of operators in the zero particle sector on the flat space with and without a boundary. The significance of the thermal expectation value of the operator in the zero particle sector is that these contribute to the one-point function of higher spin currents. After this, we extend the analysis to spheres.
We computed the partition function and Green's function on spheres and hemispheres in two and three dimensions. The partition function on a hemisphere in two and three dimensions is the sum of half of the partition function on the sphere and the corrections term, which may be interpreted as the partition function of the degrees of freedom localized on the boundary. 
We studied the large radius of expansion of the partition function and the Green's function. This will give the finite size corrections to various thermodynamics functions.
We also computed the partition function on a squashed three-dimensional sphere and studied it in the expansion of squashing parameters. 

There are a few directions which we leave for the future exploration:

We have restricted ourselves to free theories. The most interesting situation is to compute the thermodynamics of an interacting non-relativistic system on a generic background. For example, it would be interesting to evaluate the partition function for the unitary fermi system and spin-orbit coupled bosonic system on a generic curved background. This will provide a generic phase diagram as a function of the background potentials. We hope to come back to this in the near future~\cite{Meenu2025}.

Another interesting question to explore is the computation of the thermal partition function in a supersymmetric theory. Supersymmetric theories consist of both bosonic and fermionic degrees of freedom together with interactions respecting supersymmetry. As we have noted in the presence of a generic background, the bosonic and fermionic contributions to the partition functions need not cancel.  It would be interesting to investigate the partition function and the supersymmetric index in non-relativistic supersymmetric theories, especially the phase diagram and walls of discontinuity in the case of the index.
\section*{Acknowledgments}
The work of R Gupta is supported by SERB MATRICS grant MTR/2022/000291 and CRG/2023/001388. 
Meenu would like to thank the Council of Scientific and Industrial Research (CSIR), Government of India, for the financial support through a research fellowship (Award No.09/1005(0038)/2020-EMR-I). R Gupta would also like to thank Sameer Murthy and especially Justin R. David for the useful discussion and explaining some parts of his work in~\cite{David:2024pir}.

\appendix


\providecommand{\href}[2]{#2}\begingroup\raggedright\endgroup
\end{document}